\documentclass[prl,twocolumn,showpacs,preprintnumbers,amsmath,amssymb]{revtex4}
\usepackage{graphicx}
\usepackage{amsmath}
\usepackage{amsbsy}
\usepackage{dcolumn}
\usepackage{bm}

\begin{document}

\newcommand{\w}{\omega}
\newcommand{\sign}{\text{sign}}
\newcommand{\s}{\sigma}
\newcommand{\RE}{\text{Re}}
\newcommand{\IM}{\text{Im}}
\newcommand{\g}{\gamma}
\newcommand{\G}{\Gamma}
\newcommand{\Nf}{N_0}
\newcommand{\pf}{PF}
\newcommand{\be}{\begin{equation}}
\newcommand{\bea}{\begin{eqnarray}}
\newcommand{\ee}{\end{equation}}
\newcommand{\eea}{\end{eqnarray}}
\newcommand{\bwt}{\begin{widetext}}
\newcommand{\ewt}{\end{widetext}}
\renewcommand{\vec}[1]{{\mathbf #1}}


\title{
Non-Equilibrium Transport through a Kondo-Dot in a Magnetic
Field: \\
Perturbation Theory and Poor Man's Scaling}

\author{A. Rosch, J. Paaske, J. Kroha and P. W\"olfle}

\affiliation{Institut f\"ur Theorie 
der Kondensierten Materie, Universit\"at Karlsruhe, 
D-76128 Karlsruhe.}

\date{\today}

\begin{abstract}
We consider electron transport through a quantum dot described by the
Kondo model in the regime of large transport voltage $V$ in
the presence of a magnetic field $B$ with $\max(V,B)\gg T_K$.  
The electric current $I$ and the local magnetization $M$ are found to
be universal functions of $V/T_K$ and $B/T_K$, where $T_K$ is the
equilibrium Kondo temperature.  We present a generalization of the
perturbative renormalization group to frequency dependent coupling
functions, as necessitated by the structure of bare perturbation
theory.  We calculate $I$ and $M$ within a poor man's scaling approach
and find excellent agreement with experiment.

\end{abstract}


\maketitle


The transport of electrons through quantum dots  in the Coulomb
blockade regime is strongly suppressed.
In the case of an odd number of electrons on the dot,
however, when the dot carries a spin $\vec S$, exchange
coupling of $\vec{S}$ to the spin of the conduction electrons in the two
leads $(L,R)$ gives rise to a Kondo resonance at low temperature $T
\ll T_K$ ($T_K$ is the Kondo temperature).  Electrons may then cross
the dot via resonance tunneling, thus circumventing the Coulomb
blockade, as seen in a number of recent experiments
\cite{Goldhaber98,Ralph94}. 
While there is a wealth of methods like Bethe ansatz, conformal field
theory,  numerical renormalization group (RG) and self-consistent
perturbation theory available to treat the Kondo model in
equilibrium, most of these methods fail in the presence of a finite
bias voltage $V$, for $V \geq T_K$.

The first theoretical treatment of this problem dates back to the
1960s \cite{Oldpert} when the current through a Kondo impurity, in
the presence of finite $V$ and magnetic field $B$, was calculated in
perturbation theory (PT) including leading logarithmic corrections.
These early works also attempted a resummation of the logarithms 
within Nagaoka's or Abrikosov's approximations. All of these works,
however, neglected salient non-equilibrium physics of this problem
(see below).

A crucial difference between a Kondo dot in the regime of large
voltage $V \gg T_K$ and the equilibrium Kondo problem is the presence
of inelastic processes, associated with the finite current through the
dot, down to the lowest temperatures \cite{Meir94,Rosch01}.  These
processes destroy coherence on an energy scale $\Gamma$ and prevent
the full formation of the Kondo singlet resonance state.  As discussed
in \cite{Rosch01}, for a conventional Kondo dot described by an
Anderson model, $\Gamma \sim V/\ln^2(V/T_K)\gg T_{K}$ for $V\gg T_{K}$.
In this regime, which we consider here, a perturbative treatment is
sufficient to capture the developing Kondo correlations.  As is well
known from the usual Kondo effect, even in the perturbative regime it
is necessary to resum the leading logarithmic terms in order to recover
the scaling behavior.
For the equilibrium case it is known how to
achieve this in a systematic and controlled way by employing the
perturbative renormalization group method.  For the nonequilibrium
Kondo problem a comparable treatment has not been developed so far. A
certain resummation of perturbation theory (PT) has been considered
\cite{Konig96} within an Anderson model for the differential
conductance $G(V,B)$. However, the summation of leading logarithms
remained incomplete.  

In this Letter we present the first systematic study of renormalized
PT for the nonequilibrium Kondo problem,
considering in particular the influence of a magnetic field $B$. We
focus on the regime $V \gg T_K$, for any $B$. Our results are also
applicable for large $B \gg T_K$, and arbitrary $V$. In both cases the full
formation of the Kondo resonance is inhibited, and a
perturbative treatment in the exchange coupling $J$ is possible.
At first sight one might expect the nonequilibrium Kondo effect to be
similar to the high temperature regime $T\gg T_K$ of the usual Kondo
model.  It differs, however, in three qualitative aspects from the latter:
(1) The occupation probabilities of
the local spin are not thermal and have to be determined by solving
a quantum Boltzmann equation. This leads to an unusual dependence e.g.
of the spin susceptibility on $V$ and to a novel structure
of the logarithmic corrections in PT.
(2) Electrons in a wide range of energies $\sim V$ contribute to the low-$T$
properties. In contrast to the case of large $T$, where all features
are smeared on the scale $T$, we find that it is therefore essential to keep
track of the frequency dependence of the renormalized coupling
constants.  (3) Life-time effects play a
crucial role in non-equilibrium, as the corresponding rates are larger
than $T$ but smaller than $V$ \cite{Rosch01}. 

As a first step we present and discuss the results of bare PT for the
current $I(V,B)$ and the magnetization $M(V,B)$ in leading logarithmic
order.  Next we turn to the derivation of renormalization group
equations for the running coupling constants.  The structure of bare
PT forces one to keep the frequency dependence of the coupling
functions.  Employing simplifications in the spirit of Anderson's poor
man's scaling, we derive a set of RG equations in one loop order.
These have to be supplemented by a self-consistent determination of
the relaxation rate $\Gamma$, providing a cutoff of the RG flow.
Finally, the physical quantities current I, magnetization M and
differential conductance G are calculated to leading order in
$1/\ln[\max(V,B)/T_K]$ by inserting the
renormalized frequency dependent coupling functions into the lowest
order PT expressions.  In this way, all physical quantities show
scaling behavior in $V/T_K$ and $B/T_K$, where $T_K$ is the equilibrium
Kondo temperature. 

We consider the Kondo  Hamiltonian 
\begin{eqnarray}
 H&=&\sum_{\alpha=L,R,{\bf k},\s}(\varepsilon_{{\bf k}}-\mu_{\alpha})
c^{\dagger}_{\alpha{\bf k}\s}c_{\alpha{\bf k}\s}- B S_{z}\nonumber\\
& & +\frac{1}{2} \!\!\!\sum_{\alpha,\alpha'=L,R,{\bf k},{\bf k}',\s,\s'}
\!\!\!\!\!J_{\alpha'\alpha}\,
\vec{S}\cdot (c^{\dagger}_{\alpha'{\bf k}'\s'}
{\boldsymbol \tau}_{\s'\s}c_{\alpha{\bf k}\s}), 
\end{eqnarray}
where $\mu_{L,R}=\pm V/2$. $\vec{S}$ is the spin 1/2 on the
dot and ${\boldsymbol \tau}$ are the Pauli matrices. We shall use the
dimensionless coupling constants $g_d=\Nf J_{LL}=\Nf J_{RR}$ and
$g_{LR}=\Nf J_{LR}=\Nf J_{RL}$ (assuming a symmetric dot), where $\Nf$
is the local density of states. The local spin
$\vec{S}=\frac{1}{2}\sum_{\gamma \gamma'}f^{\dagger}_{\gamma}
{\boldsymbol\tau}_{\gamma\gamma'}f_{\gamma'}$
is represented by pseudo-fermions (\pf) in the sector of the Hilbert
space with $\sum_{\gamma} f^{\dagger}_{\gamma}f_{\gamma}=1$.

\begin{figure}
\begin{center}
\includegraphics[width=0.99\linewidth]{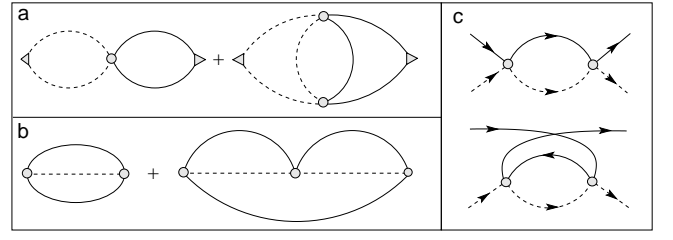}
\end{center}
\vspace*{-1.2em}
\caption{\label{fig1} Feynman diagrams for (a) $D_{LR}$, (b) \pf\
 self-energies and (c) vertices entering the 1-loop RG equation.
\pf\ (electron-) propagators are displayed as dashed  (full)  lines.}
\vspace*{-0.5cm}
\end{figure}

The electric current $I=(e/\hbar)J_{LR}\IM[D^{K}_{LR}(t,t)]$ through
the quantum dot  may be conveniently expressed through the Keldysh
component 
of the contour-ordered correlation function
$D_{LR}(t,t')=(-i)^{2}\sum_{{\bf k},{\bf k}'}
\left\langle T_{c}\left\{
c^{\dagger}_{L{\bf k}'\s'}(t)\frac{{\bf \tau}_{\s'\s}}{2}c_{R{\bf
k}\s}(t)\cdot{\bf S}(t')\right\}\right\rangle$
%
To obtain the leading logarithmic corrections, we evaluate the diagrams
shown in Fig.~\ref{fig1}a, which for $T, V, B$ much smaller 
than the band cutoff $D$ yield
\begin{eqnarray}
I&=&\frac{2 e}{2 \pi \hbar}\left(\frac{\pi}{2} g_{LR}\right)^{2}\bigl[
u_2(V)+u_2(V+ B)+u_2(V- B)\nonumber\\
&&-M (c(B)-c(-B)) \bigr], \label{Ipert}
\end{eqnarray}
with $u_p(x)=x\left(1+2 p g_d \ln \frac{D}{|x|}\right)$, ($p=1,2$),and
$c(B)=\coth\frac{V+ B}{2 T}[u_1(V+B)+2 V g_d \ln\frac{D}{|V|}+2 B g_d \ln \frac{D}{|B|}]$
where all logarithms are implicitly cutoff by $T$.
Appelbaum's expression for the current \cite{Oldpert} differs from
(\ref{Ipert}) in that he assumed $M$ to be given by
its thermal equilibrium value.
In the zero-field limit $M$ vanishes and
(\ref{Ipert}) reduces to $I=\frac{2 e}{2 \pi \hbar}(\frac{\pi}{2}
g_{LR})^{2}3V \left(1+4g_{d}\ln\frac{D}{|V|}\right)$ derived 
earlier in Refs.~\cite{Kaminski00,Oldpert}.
To calculate the current for finite $B$, one needs to know the local
magnetization $M=n_{\uparrow}-n_{\downarrow}$, and at finite voltage
this necessarily involves solving a quantum Boltzmann equation. 
In diagrammatic language this is nothing but the 
Dyson equation of the lesser \pf\
Green function $G_{\gamma}^{<}(\w)$ in steady state:
$G_{\gamma}^{<}(\w)\G_{\gamma}(\w)=\Sigma_{\gamma}^{<}(\w)A_{\gamma}(\w)$,
where $\G_\gamma$ is the imaginary part of the self-energy and $A_\gamma$ the
\pf\  spectral function. 
The leading logarithmic corrections are obtained from
the self-energy diagrams in
Fig.~\ref{fig1}b, yielding

\begin{eqnarray}
M\!&=&\frac{\cal T}{\coth\frac{B}{2T}\left[ \frac{\cal
    T}{2}+g_-^2 u_1( B)
\right]+g_{LR}^2 (c(B)+c(-B))} \label{Mpert}\\%
{\cal T}&=&2 g_+^2  u_1(B)+
2 g_{LR}^2 [u_1(V+B)-u_1(V-B)-2 B] \nonumber
 \end{eqnarray}
with $g_{\pm}^2=(g_d^2 \pm g_{LR}^2)$.
Note that the magnetization differs from the equilibrium result
even to order $g^0$, as has been derived independently in \cite{Private}
(see also Eq.~(\ref{goldenM}) below) in disagreement
with \cite{Coleman01}. Most
notably, the magnetic susceptibility changes from the usual $1/T$
Curie law to a $1/V$ behavior for $V\gg T$. Corrections of  order
$g^3 \ln D$ to the collision integrals in the quantum Boltzmann equation
 result in corrections of order $g\ln D$ to $M$. Thus, they
are much larger than the usual $g^{2}\ln D$ corrections 
obtained in  equilibrium.  In the limit  $V\to 0$,
Eq.~(\ref{Mpert}) simplifies to the non-interacting
result $M=\tanh(B/2T)$, as we have neglected the
subleading $g^{2}\ln D$-corrections which can be obtained by
including the $\RE\Sigma$ shift of the \pf\  energy levels.  The effects of
$\IM\Sigma$ are more important and are discussed below.



In the scaling regime, $V,B\ll D$, bare PT is not valid and has to be
resummed even for small couplings $g$ and in the weak-coupling regime
$V,B\gg T_K$ (see Fig.~\ref{figG}b).  The method of choice for such a resummation is the
perturbative RG, using the basic idea that a
change of the cutoff $D$ can be absorbed in a redefinition of the
coupling constants $g$ \cite{Anderson70}. 

A close inspection of Eqs.~(\ref{Ipert},\ref{Mpert}) indeed reveals
that in the regime where, under renormalization, $D$ gets smaller than
$V$, it is {\em not} possible to absorb a change of $D$ in a
redefinition of the couplings.  For example, logarithmic
corrections to $g_{LR}$ in the denominator of Eq.~(\ref{Mpert}) are
proportional to $2 \ln(D/V)$ for $B\to 0$ while the analogous
correction in the numerator takes the form $\ln(D/V) + \ln( D/T)$. This
apparent breakdown of scaling is due to the fact that electrons in a
finite energy window $\mu_R \lesssim \w \lesssim \mu_L$
contribute to low-energy properties. How much their scattering is
renormalized will depend on their respective positions within
this window: upon renormalization the coupling ``constants'' acquire
{\em frequency dependences}. Taking
\begin{figure}
\begin{center}
\includegraphics[angle=-90,width= 0.9\linewidth]{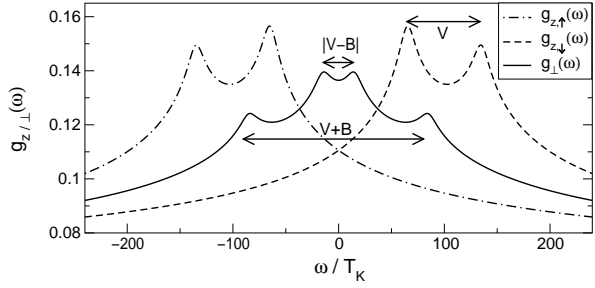}
\end{center}
\vspace*{-2.0em}
\caption{\label{figJ} Renormalized coupling constants $\tilde{g}_{z,\sigma}(\w)$ 
(dashed lines) and $\tilde{g}_{\perp}(\w)$ (solid line) for $B=100 T_K$ and
$V=70 T_K$. For these values  
Eqs.\ (\ref{jRG}, \ref{goldenG}) yield $\Gamma = 6.47 T_K$.}
\vspace*{-0.5cm}
\label{fig2}
\end{figure}
this properly into account, one recovers scaling. One way to
derive perturbative RG equations in such a situation is to start from
so-called ``exact'' RG equations for  ($\w$-dependent)
one-particle irreducible Green functions by
a straightforward
generalization of equilibrium RG
methods \cite{Salmhofer01}.  We will
not follow this route here, but suggest a substantially simpler (but
less systematic) approach.  A considerably more involved real-time RG
scheme has been proposed by
Schoeller and K\"onig \cite{Schoeller00},
but has not yet been applied to the present problem.

We start from the observation, that all logarithmic terms in 
next-to-leading order PT stem from the simple vertex
renormalizations shown in Fig.~\ref{fig1}c when the real part $\approx
1/(\w \pm B/2)$ of the \pf\ Green function is convoluted with the Keldysh
component of the electron line $ - 2 \pi i \Nf \tanh[(\w-\mu_\alpha)/2T]$.
Using cutoffs symmetric with respect to $\mu_{L,R}$, respectively, one
obtains at $T=0$
\begin{eqnarray}\label{logDer}
\frac{\partial}{\partial \ln D}
\int_{-D}^D d\w \frac{\sign \ \w}{\w-\Delta \w}\approx 2
\Theta(D-|\Delta \w|),
\end{eqnarray}
where $\Delta \w$ depends on $\mu_\alpha$, $B$ and the incoming and
outgoing frequencies. Now we include the effect of frequency and
spin-dependent coupling constants denoted by $g^{\alpha\sigma\w_c;
\alpha'\sigma'\w_c'}_{\gamma\w_f; \gamma'\w_f'}$ for an incoming
electron in lead $\alpha=L,R$ with energy $\w_c$ and spin
$\sigma=\uparrow,\downarrow$ interacting with a \pf\ of spin $\gamma$ and 
frequency $\w_f$ describing the local spin.  Primed
quantities refer to  outgoing particles. Generally,
vertices are complex and have Keldysh indices, but we will only keep track
of the real parts  on one Keldysh contour, which we believe
to be justified to leading order in 
$1/\ln[V/T_K]$. The pseudo fermion spectral functions are strongly peaked
at $\omega=\pm B/2$, which allows setting $\omega_f (\omega{'}_f)$ to
$-\gamma\frac{B}{2} (-\gamma{'}\frac{B}{2})$.  Furthermore,
neglecting the remaining frequency
dependence of the coupling functions in the frequency window
$|\omega|<D$ \cite{Valid}, one is led \cite{Unpubl} to two RG equations
for the coupling functions $\tilde{g}_{z\sigma}(\omega)$ and 
$\tilde{g}_{\perp}(\omega)$
(assuming $J_{LL}=J_{RR}=J_{LR}=J_{RL}=J$)
\begin{eqnarray}
\frac{\partial \tilde{g}_{z\sigma}(\w)}{\partial \ln D} &=&
-\sum_{\beta=-1,1} \tilde{g}_\perp(\frac{B+\beta V}{2})^2 
\Theta_{\w+\sigma (B+ \beta \frac{V}{2})}
 \label{jRG} \\
\frac{\partial \tilde{g}_{\perp}(\w)}{\partial \ln D} &=&
-\!\!\! \sum_{\beta,\sigma=-1,1} \!\!\! \frac{\tilde{g}_\perp(\frac{\sigma B+\beta V}{2})
\tilde{g}_{z\sigma}(\frac{\beta V}{2})}{2}
\Theta_{\w+\frac{\sigma B+\beta V}{2}} \nonumber
\end{eqnarray}
with $\tilde{g}_{\perp}(\w)=\tilde{g}_{\perp}(-\w)$, $\tilde{g}_{z\uparrow}(\w)=\tilde{g}_{z\downarrow}(-\w)$ and \pagebreak
\begin{eqnarray}
g^{\alpha \sigma,\w;\alpha' \sigma,\w}_{\gamma,-\gamma B/2;
\gamma,-\gamma B/2}&=&\tau^z_{\gamma \gamma} \tau^z_{\sigma \sigma}
\tilde{g}_{z\sigma}(\w)\\
g^{\alpha \sigma, \w;\alpha' \bar{\sigma},\w- \gamma B}_{\gamma,-\gamma B/2;
\bar{\gamma},-\bar{\gamma} B/2}
&=&(\tau^x_{\gamma \bar{\gamma}} \tau^x_{\sigma \bar{\sigma}}+
\tau^y_{\gamma \bar{\gamma}} \tau^y_{\sigma \bar{\sigma}}) \, 
\tilde{g}_{\perp}(\w- \gamma B/2)\nonumber,
\end{eqnarray}
with the initial conditions $\tilde g_{z\sigma}(\omega) = \tilde
g_{\perp}(\omega) = JN_0$ at $D = D_0$, the bare cutoff, and
$\Theta_{\Delta\w}=\Theta(D-|\Delta\w|)$.
Despite the fact that we use the same functions for both diagonal
(LL,RR) and off-diagonal (LR,RL) processes, their renormalization is
drastically different as  different frequency ranges  are probed in
the two cases (see below). In various limits analytic solutions are
possible but as the formulas turn out to be rather lengthy, we will
here restrict ourselves mainly to numerical results.

The RG equations (\ref{jRG}) are valid in the regime of cutoff values
$D$ where the effect of spin-flip processes destroying the quantum
coherence of the Kondo bound-state is negligible, or $D >\Gamma$.
Due to the finite current, relaxation processes contributing to
$\Gamma$ are present even at $T=0$, and lead to an imaginary
part of the self-energy of the pseudofermions (e.g. in the PF
propagator in the diagrams of Fig.~\ref{fig1}c) and to vertex
corrections. Although slightly different relaxation rates may enter
in the various coupling constants, with logarithmic accuracy we may
use a single rate $\Gamma$. As all relevant processes involve at
least one spin-flip, we identify $\Gamma$ with the transverse spin
relaxation rate $1/T_2$, which is given in terms of the renormalized
couplings as
\begin{eqnarray}
\Gamma&=&  \frac{\pi}{4 \hbar} \sum_{\substack{\alpha,\alpha'=L,R\\
 \gamma=-1,1}}
\int \!\! d\w \Bigl[\tilde{g}_{z\gamma}(\w)^2
f_{\w-\mu_\alpha}(1-f_{\w-\mu_{\alpha'}})\label{goldenG}\\[-.2cm] 
&&  \hspace{1.4cm}   +\tilde{g}_\perp(\w-\gamma B/2)^2
f_{\w-\mu_{\alpha}}(1-f_{\w-\mu_{\alpha'}- \gamma B})\Bigr].
\nonumber
\end{eqnarray}
Note the missing factor $2$ in front of the spin-flip term
$\tilde{g}_{\perp}(\w-\gamma B/2)^2$ in $\Gamma$ (compared to the
corresponding term in (\ref{goldenI}), below), which reflects the fact that
$1/T_2$ \cite{Woelfle71} is {\em not} given by a 
pseudo-Fermion self-energy and arises from vertex corrections in a
linear-response calculation of the transverse susceptibility. 
The effect of $T_2$ is to provide an upper time
cutoff for the coherent scattering processes entering Eq.\ (\ref{jRG}).
Hence, it is modeled phenomenologically by replacing
$\Theta_\omega$ with $\Theta (D-\sqrt{\omega^2 + \Gamma^2})$ in
Eq.\ (\ref{jRG}).

A self-consistent solution of Eqs. (\ref{jRG}, \ref{goldenG}) can now
be obtained with little numerical effort (formally, effects of
self-consistency for $\Gamma$ are subleading for $V,B\gg T_K$).  The
resulting couplings $\tilde{g}_{z/\perp}(\w)$ are shown in
Fig.~\ref{figJ}. In the renormalization process the $\tilde
g_{z/\perp}(\omega)$ develop peaks at frequencies $\omega = -\sigma
B\pm\frac{V}{2}$ for $\tilde g_{z\sigma}$ and $\omega = \pm
\frac{1}{2} (B\pm V)$ for $\tilde g_{\perp}$, at which resonant
scattering from one Fermi surface to another becomes possible. Only
these resonant processes survive for small running cutoff $D$, as can
be seen from the structure of the functions
$\Theta_{\w}=\Theta(D-|\w|)$ in (\ref{jRG}). This Kondo-type resonant
enhancement is eventually cutoff by $\Gamma$.  Note that $\Gamma$
remains larger than $T_K$ for $V,B > T_K$ prohibiting the
flow towards strong coupling contrary to the suggestion of
\cite{Coleman01} (Ref.~\cite{Rosch01} discusses conditions for
strong-coupling physics in the regime $V\gg T_K$).

\begin{figure}
\begin{center}
\includegraphics[angle=-90,width=\linewidth]{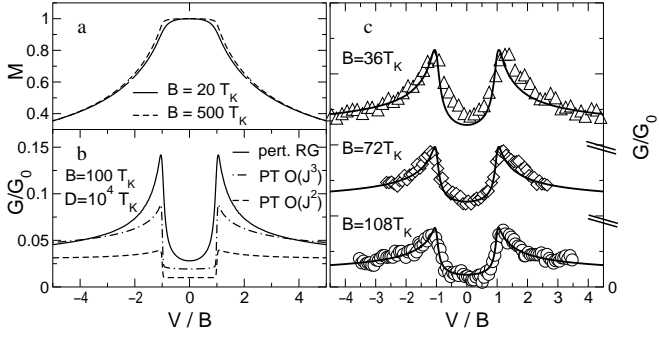}
\end{center}
 \vspace*{-.3cm}
\caption{\label{figG}
a)Local magnetization $M(V/B,B/T_K)$ of a sym\-met\-ric dot for fixed
  magnetic field $B$.  
b)Differential conductance in units of $G_0=\frac{e^2}{\pi \hbar}$. 
  Comparison of perturbative RG and bare PT to $O(J^2)$ and $O(J^3)$.
c)Conductance measurements 
  of Ref.\ \cite{Ralph94} (symbols) on metallic point contacts
  in magnetic fields $0.85,1.7,2.55\,$T ($B=36,72,104\, T_K$,  with
  $T_K\approx30$mK \cite{Ralph94}).  Assuming that the corresponding
  point contact is described by a single-channel ($J_{LR}=\sqrt{J_{LL}
    J_{RR}}$) Kondo model, $\frac{J_{RR}}{J_{LL}}\approx
  4.2$ is determined from $G(V\!\!=\!\!0,B\!\!=\!\!0,T\!\!=\!\!50
  \text{mK})=\frac{4 J_{LL} J_{RR}}{(J_{LL}+J_{RR})^2}
  G_{\text{sym}}(T/T_K)$, where $G_{\text{sym}}$ is known exactly from
  NRG calculations \cite{Costi94}.  This fixes {\em all} parameters for
  our RG calculation (solid lines, $T=0$) which uses a straightforward
  generalization of (\ref{jRG}--\ref{goldenM}) for $J_{LL}\neq J_{RR}$. 
  As the ($B$-dependent) background  is not known experimentally,
  we subtract $\Delta G=G_B-5.2\cdot 10^{-5} G_0 \frac{V}{T_K}$, where
  $G_B$ is fitted to our results at large $V$. 
  The experimental temperature 
  $T=50$mK leads to an extra small broadening at $V=B$.}
\vspace*{-0.5cm}
\end{figure}
Knowing the renormalized $\tilde{g}_{z/\perp}(\w)$ at $D=0$, the 
magnetization $M$ can be found from solving the Boltzmann equation
$\partial_t n_\sigma = 0$ or
\begin{eqnarray}
&& \hspace*{-.5cm} n_\uparrow \!\!\!\! \sum_{\alpha,\alpha'=L,R} 
\int \! d\w \, \tilde{g}_{\perp}(\w- B/2)^2
f_{\w-\mu_\alpha}(1-f_{\w-\mu_{\alpha'}- B})  \label{goldenM}\\
&& \hspace*{-.5cm} =
n_\downarrow  \!\!\!\! \sum_{\alpha,\alpha'=L,R} \int\!  d\w \,  
\tilde{g}_{\perp}(\w+ B/2)^2 
f_{\w-\mu_\alpha}(1-f_{\w-\mu_{\alpha'}+B}),\nonumber 
\end{eqnarray}
the solution of which in turn determines the current as
\begin{eqnarray}
I&=& \frac{ \pi e}{16 \hbar} \int \!\! d\w 
\sum_{\sigma=-1,1}\Bigl[
\tilde{g}_{z\sigma}(\w)^2 f_{\w-\mu_L}(1-f_{\w-\mu_R})
  \label{goldenI}\\
 &&\hspace*{-0.9cm}+ 4 \tilde{g}_{\perp}(\w-\sigma B/2)^2  
f_{\w-\mu_L}(1-f_{\w-\mu_R- \sigma B}) n_\sigma 
 \Bigr] - 
(L \leftrightarrow R),\nonumber
\end{eqnarray}
In Eqs.~(\ref{goldenG}--\ref{goldenI}) there are further lifetime
effects, which become relevant for $|V-B|,T \lesssim \Gamma$.
This broadening we model phenomenologically by smearing the Fermi
functions over a width $\Gamma$ \cite{Unpubl}. Note that expanding
these results in bare PT, one recovers Eqs.\ (\ref{Ipert},\ref{Mpert}).

In Fig.~\ref{figG} we show $M(V/B,B/T_K)$ and $G(V/B,B/T_K)$.
Within PT, $G$ shows threshold behavior at $|V|=B$, due to the opening of 
another transport channel involving spin flip, requiring the Zeeman energy $B$. 
Higher order processes enhance the steps into peaks (Fig.~\ref{figG}b).
The agreement with experiment \cite{Ralph94} 
is excellent (Fig.~\ref{figG}c), 
considering that there is no free parameter and
our result is valid up to terms of order $1/\ln[\max(V,B)/T_K]$ only. 
We emphasize that neither perturbation theory (Fig.~\ref{figG}b) nor
Appelbaum's result (as shown in \cite{Ralph94}), nor the result of
\cite{Konig96} can describe the experiment. 

In conclusion, we propose a simple method to generalize
poor-man's scaling (one-loop order) to non-equilibrium.
Compared to the equilibrium situation, it is necessary to include
non-equilibrium distribution functions, decay-rates and the
$\w$-dependence of coupling constants.  The different structures
arising even in the perturbative regime may be best exemplified by
considering the local magnetic  susceptibility for $V\gg T_K$, $B\to0$
\begin{eqnarray}
\chi(V)=\frac{2}{V}
\frac{1+2 \alpha \ln \frac{V}{T_K} (1+ \alpha \ln \frac{V}{T_K})}
{\left( 1-
 \frac{(1 + 2 \alpha \ln \frac{V}{T_K}) \ln\left[ \ln \frac{V}{T_K}
(1+\alpha  \ln \frac{V}{T_K})\right]   }{(1 +  \alpha \ln \frac{V}{T_K})
\ln  \frac{V}{T_K}} \right)^2}
\end{eqnarray}
with $\alpha=(g_d^2-g_{LR}^2)/(2 g_{LR})$ and $T_K=D e^{-1/(g_d+g_{LR})}$.
To the same order of approximation,
one obtains in equilibrium for $T\gg T_K$ just $\chi=1/T$.

We thank S.~De~Franceschi, J. K\"onig, O. Parcollet, 
H. Schoeller and 
A.~Shnirman for helpful discussions and especially L.~Glazman, who
suggested to consider the case of finite $B$. This
work was supported in part by the CFN
and the Emmy Noether program (A.R.) of the DFG.


\begin{thebibliography}{0}
\bibitem{Valid} These approximations should be valid 
to leading order in $1/\ln[V/T_K]$, as frequency
dependencies are only logarithmic or power-laws with exponents
of order $1/\ln[V/T_K]$.
\end{thebibliography}
\end{document}